\documentclass[12pt,english]{article}
\usepackage[T1]{fontenc}
\usepackage[latin9]{inputenc}
\usepackage[letterpaper]{geometry}
\usepackage{array}
\usepackage{longtable}
\usepackage{amsmath}
\usepackage{amssymb}
\usepackage{esint}
\usepackage{hyperref} %%%Hyperlinked references enviroment%%%
\usepackage{graphicx} %%%Enviroment for including figures using "includegraphics"%%%

\makeatletter

\providecommand{\tabularnewline}{\\}

\makeatother

\makeatother

\usepackage{babel}

\begin{document}
		
\title{
\hfill{\small \tt WIS/8/11-SEP-DPPA}\vskip 50pt
On $O(N_c)$ $d=3$ ${\cal N}=2$ supersymmetric QCD Theories}
\author{Ofer Aharony and Itamar Shamir\\\\
{\it Department of Particle Physics and Astrophysics}\\
{\it Weizmann Institute of Science, Rehovot 76100, Israel}\\
{\small{\tt e-mails~:~Ofer.Aharony,~Itamar.Shamir@weizmann.ac.il}}}

\maketitle

\begin{abstract}
  We study three dimensional ${\cal N}=2$ supersymmetric QCD theories with $O(N_c)$ gauge groups
  and with $N_f$ chiral multiplets in the vector representation. We argue that for $N_f < N_c-2$
  there is a runaway potential on the moduli space and no vacuum. For $N_f \geq N_c-2$ there is a
  moduli space also in the quantum theory, and for $N_f \geq N_c-1$ there is a superconformal fixed
  point at the origin of this moduli space that has a dual description as the low-energy fixed point
  of an $O(N_f-N_c+2)$ gauge theory. We test this duality in various ways; in some cases the duality
  for an $O(2)$ gauge theory may be related to the known duality for $U(1)$ gauge theories. We also discuss real mass deformations, which allow to connect theories with a different Chern-Simons level. This allows us to connect our duality with the known duality in $O(N_c)$ theories with a Chern-Simons term of level $k$, where the dual gauge group is given by $O(N_f+|k|-N_c+2)$.
\end{abstract}

\newpage

\tableofcontents

\section{Introduction}

Supersymmetry (SUSY) is a useful tool for studying the dynamics of strongly coupled gauge
theories in various dimensions. The dynamics of ${\cal N}=1$ supersymmetric QCD (SQCD) in four dimensions,
for all classical gauge groups, was understood, following the work of Seiberg, in the 1990s \cite{Seiberg:1994bz,Seiberg:1994pq,Intriligator:1995id,Intriligator:1995ne}.
These theories show some similarities and some differences from the non-supersymmetric $SU(3)$
QCD theory which is relevant for describing nature. In the pure gauge theory they exhibit
discrete vacua, as in the non-supersymmetric case. For a small number of flavors they
exhibit a runaway behavior on the moduli space, which is not present in the non-supersymmetric
case which lacks scalars. For a larger number of flavors effects like chiral symmetry breaking
and confinement, somewhat similar to those of the non-supersymmetric case, were discovered. For an
even larger number of flavors (but still in the asymptotically free regime) the theory was
found to flow (at the origin of its moduli space) to a non-trivial fixed point, and it was found
that this fixed point has an
alternative dual description as the low-energy limit of a different gauge theory, a phenomenon
known as Seiberg duality \cite{Seiberg:1994pq}. Such dualities have since been found in many $d=4$ ${\cal N}=1$ supersymmetric theories,
but their general description or derivation is not known (see \cite{Komargodski:2010mc} for an interesting attempt), and it is not clear yet if they can be applied also
in non-supersymmetric cases or not.

The dynamics of ${\cal N}=2$ supersymmetric QCD in three dimensions has been less studied, since the
motivations for studying it are less direct; some of these theories may have applications in
condensed matter physics, and some of them exhibit a Seiberg duality similar to the four
dimensional case, which should help to shed some light on the origin of Seiberg duality
(in particular, it makes it less likely that this duality is fundamentally related to
electric-magnetic duality).
%$d=3$ ${\cal N}=2$ also exhibit another duality known as mirror
%symmetry \cite{Intriligator:1996ex,deBoer:1997kr,Aharony:1997bx}, whose relation to Seiberg duality is unclear.
Flowing between four dimensional and three dimensional QCD-like theories, via a
circle compactification, has been used for a better understanding of confinement (see, for instance, \cite{Shifman:2008ja}), and studying
supersymmetric examples may be useful for understanding this better as well.

For the $d=3$ ${\cal N}=2$ SQCD theories with gauge groups $U(N_c)$ and $USp(2N_c)$, the quantum
dynamics has been understood already in 1997 \cite{Aharony:1997bx,Karch:1997ux,Aharony:1997gp}, and a picture rather similar to the four dimensional case has
emerged. For small numbers of flavors (now including also the pure gauge theory \cite{Affleck:1982as}) there is a
runaway behavior on the moduli space, while for a larger number of flavors there is a moduli space of
quantum vacua. For a large enough number of flavors (note that now this number is not limited by
asymptotic freedom) there is a conformal field theory at the origin of the moduli space, and a
dual theory was found that flows to the same conformal field theory, as in four dimensional Seiberg
duality. The dynamics of the group $SU(N_c)$ seems to be more complicated (note that $U(1)$ gauge
theories are strongly interacting at low energies in three dimensions), and so far no dual
description was found for this case.

The four dimensional Seiberg dualities may be motivated by a brane construction \cite{Elitzur:1997fh} (see \cite{Giveon:1998sr} for a review), for all
the classical gauge groups. As mentioned already in \cite{Aharony:1997gp}, the same brane construction suggests a Seiberg duality also in the
three dimensional ${\cal N}=2$ case, for the gauge groups $U(N_c)$ (with a dual gauge group $U(N_f-N_c)$),
$USp(2N_c)$ (with a dual gauge group $USp(2N_f-2N_c-2)$ and $O(N_c)$ (with a dual gauge group
$O(N_f-N_c+2)$), but not $SU(N_c)$. The quantum dynamics of the first two cases was studied in
detail in \cite{Aharony:1997gp}, and a duality similar to the one suggested by the brane construction
was found; the duality in this case involves also variables (complexified Coulomb branch coordinates) which are
not visible directly in the brane construction, so one cannot directly use the brane construction to
derive the precise form of the dual theory. A detailed test of this duality by partition function computations was recently performed in \cite{Willett:2011gp}. The dynamics of the $O(N_c)$ case has not yet been
studied, and presents some subtleties that are not present in the other cases (as in four dimensions \cite{Intriligator:1995id}).
A natural generalization of QCD in three dimensions is the addition of a Chern-Simons (CS) term, and the
duality symmetries can be generalized to this case as well \cite{Giveon:2008zn}; the ${\cal N}=2$
theory with a CS term is
simpler since it lacks a Coulomb branch. Dualities for the theory with the CS term can be derived from the SQCD case by
adding real masses to some flavors in SQCD theories with no CS term, generating an effective CS term at low energies.
This generalization has also been studied for the $O(N_c)$ case in \cite{Kapustin:2011gh}. Flows between theories with
different numbers of colors, flavors and CS levels were studied in detail in \cite{ItamarThesis} for all the classical gauge groups (including the $O(N_c)$ case).

In this paper we study the quantum dynamics of the $d=3$ ${\cal N}=2$ SQCD theory with $O(N_c)$ gauge group and $N_f$ chiral multiplets in the vector representation,
and the corresponding dualities. The picture we find is very similar to the one found for $U(N_c)$ and
$USp(2N_c)$ gauge groups, as suggested by the brane construction. In particular, as in the $U(N_c)$ case,
we will find a simple picture for $O(N_c)$ gauge groups, but not (in general) for $SO(N_c)$ gauge groups.
The differences between the $d=3$ and $d=4$ dualities are also the same for the $O(N_c)$ case as for
other cases; also here there is no case in which the dual theory is free at low energies (unlike the
$d=4$ case), and also here there is no known description of the dual theory at high energies (since
the superpotential of the dual theory involves the effective Coulomb branch coordinate), so it should be treated as
an effective theory defined below some scale.

We begin in section 2 with a review of some general properties of $d=3$ ${\cal N}=2$ theories with
$O(N_c)$ gauge groups. In section \ref{duality} we discuss the duality we suggest for $N_f \geq N_c-1$ and some
of its tests. In section 4 we describe the flow to smaller numbers of flavors and the resulting
dynamics. In section 5 we discuss flows by real masses, which allow us to connect our duality to the
duality of \cite{Kapustin:2011gh} in the presence of a Chern-Simons term. We end in section 6 with a discussion and summary.

{\bf Note added} : As this paper was being completed, two papers appeared \cite{Benini:2011mf,Hwang:2011ht} which also suggest the
same duality for the $O(N_c)$ theory. These papers do not test the duality in the same ways that
we do, but instead they test it by computing the partition function and supersymmetric indices of
the two dual theories, finding a precise match. Thus, these papers provide complementary evidence
for the validity of this duality.

\section{Generalities of $d=3$ $O(N_c)$ ${\cal N}=2$ SUSY gauge theories}

In this section we review some necessary background. The quickest approach to classical $d=3$ $\mathcal{N}=2$ SUSY theories is by dimensional reduction of the more familiar $\mathcal{N}=1$ SUSY theories in $d=4$. The reduced theory inherits the holomorphic properties of $d=4$ $\mathcal{N}=1$ theories, providing strong control over the dynamics, yet it allows a richer set of deformations and moduli spaces. We emphasize the unique aspects of the $d=3$ theory.

For many purposes it does not matter if the gauge group is $O(N_c)$ or $SO(N_c)$, but in the next sections
we will find that the quantum dynamics is simpler for the $O(N_c)$ case, so we will focus on this
case here, and only sometimes discuss also the $SO(N_c)$ case.
We shall call the transformations with determinant $(-1)$, which distinguish between the special and the non-special orthogonal group, reflections (not to be confused with space time parity).

\subsection{Action of $d=3$ $\mathcal{N}=2$ SUSY gauge theories}

The vector multiplet of $\mathcal{N}=2$ SUSY in $d=3$, in Wess-Zumino gauge, contains a vector field $A_{\mu}$, a real scalar $\phi$, a fermion $\lambda$ and an auxiliary field $D$, all in the adjoint representation. The Lagrangian is readily obtained by dimensional reduction from $d=4$,
\begin{align}
\frac{1}{4 g^2} \mathrm{Re} \left[ \int \mathrm{d}^{2} \theta \mathrm{Tr} (W^{\alpha}W_{\alpha})
 \right] = & \frac{1}{2 g^2}\mathrm{Tr}\left(-\frac{1}{4}F_{\mu\nu}F^{\mu\nu} - \frac{1}{2}\mathcal{D^{\mu}}\phi\mathcal{D_{\mu}}\phi - i\bar{\lambda}\bar{\sigma}^{\mu}\mathcal{D}_{\mu}{\lambda} - \phi [\lambda,\bar{\lambda}] + \frac{1}{2}DD\right),
\label{eq:action for gauge field}
 \end{align}
where the generators are normalized to $\mathrm{Tr}(T^{a}T^{b})=2\delta^{ab}$. The flat directions obtained when the real scalars above assume expectation values are a new feature of the $d=3$ theory.

Note the fermion contraction in the second term from the right in \eqref{eq:action for gauge field}, which has the structure of $\lambda \bar{\lambda}= i \lambda^\alpha \epsilon_{\alpha \beta} \bar{\lambda}^\beta$. This is Lorentz-invariant since in $d=3$ there is no distinction between dotted and undotted spinor indices. When $\phi$ has a vacuum expectation value this term gives a mass to the fermion. In general, a fermion mass term of the form $\psi \bar{\psi}$ is called a real mass, since the mass parameter is fixed to be real. This term is distinct from the mass term $(m \psi \psi + c.c.)$ familiar from $d=4$, which in this context is called a complex mass.

We also consider matter fields in the fundamental (vector) representation, given by chiral superfields $Q$ (the component fields are just as in $d=4$; we denote also the scalar component of the chiral
superfield by $Q$, hoping that this causes no confusion). The classical action is
\begin{align}
\int\mathrm{d}^{4}\theta Q^{\dagger}e^{2V} Q \nonumber
 = & -\mathcal{D}^{m}Q^{*}\mathcal{D}_{m}Q - \left|\phi Q\right|^{2} - \frac{i}{2} \psi\sigma^{m}\mathcal{D}_{m} \bar{\psi} + \frac{i}{2}\mathcal{D}_{m}\psi\sigma^{m}\bar{\psi} - \psi\phi\bar{\psi}\nonumber \\
   & + F^{*}F + i\sqrt{2}\left(Q^{*}\lambda\psi-Q\bar{\lambda}\bar{\psi}\right) + Q^{*}DQ.
\label{eq:action for csf}
\end{align}
Taking $N_f$ matter fields $Q^i$ we have the global symmetry group $SU(N_f)\times U(1)_{R}\times U(1)_{A}$, with quantum numbers $W_{\alpha}:(\mathbf{1},1,0)$ and $Q:(\mathbf{N_f},0,1)$.

We can introduce real masses for the chiral superfields by deforming the K\"ahler potential to
\begin{equation}
\int\mathrm{d}^{4}\theta Q^{\dagger}e^{2V} e^{-2 m_R \theta \bar{\theta}} Q = - m_R^2 |Q|^2 - m_R \psi \bar{\psi} +  \cdots.
\label{realmass}
\end{equation}
It is apparent that $m_R$ enters \eqref{realmass} exactly as the vacuum expectation value of $\phi$ enters \eqref{eq:action for csf}. In fact, we can allow $m_R$ to be non-trivial in flavor space. It is then natural to interpret $m_R$ as the scalar in a background gauge field for the flavor symmetry $SU(N_f)\times U(1)_A$.

\subsection{The classical moduli space}

Since we have not considered a tree-level superpotential so far, there
is only a $D$-term potential
\begin{equation}
V=g^{2}\left(Q^{\dagger}T^{a}Q\right)^{2}+\left|\phi Q\right|^{2}.
\label{eq:so(N_c) scalar potential}
\end{equation}
Consider first zeros of the potential obeying $\phi=0$. In this case, up to gauge and global symmetry transformations, as in $d=4$, we can write
\begin{equation}
Q=\left( \begin{array}{ccccc}
v_{1}\\
 & \ddots\\
 &  & v_{N_f} & &
\end{array}
\right)
\end{equation}
for $N_f<N_c$, and if $N_f\geq N_c$
\begin{equation}
Q=\left(\begin{array}{ccc}
v_{1}\\
 & \ddots\\
 &  & v_{N_c} \\
 &
\end{array}
\right),
\end{equation}
with $v_{j} \geq 0$ in both cases (for $O(N_c)$ gauge theories). In the generic case where all the $v_i \neq 0$, the gauge symmetry is broken to $O(N_c-N_f)$ if $N_f<N_c$, and for $N_f \geq N_c$ it is completely broken. This part of the moduli space is called the Higgs branch.

The Higgs branch can also be described by gauge-invariant meson superfields $M^{ij} = Q_a^i Q_a^j$, in a
symmetric tensor representation of $SU(N_f)$. In the
$SO(N_c)$ case (but not for $O(N_c)$), when $N_f \geq N_c$ there are also independent baryon operators $B^{i_1 , \ldots , i_{N_c}} = Q^{i_1}_{[a_1}  \ldots  Q^{i_{N_c}}_{a_{N_c}]}$.

There are also zeros of the potential where $\phi \neq 0$. Using a gauge transformation we can always choose $\phi$ to lie in the Cartan sub-algebra. It is convenient to choose a basis for the algebra such that the Cartan sub-algebra is given by diagonal matrices; this choice for the description of the $O(N)$ group and algebra is described in appendix \ref{appA}. For $N_c=2n$ we then have
\begin{equation} \label{eq:eigenvalues phi N_c=2n}
\phi =  \mathrm{diag}(\phi_{1} , \ldots , \phi_{n} , -\phi_{1} , \ldots , -\phi_{n}),
\end{equation}
where using the Weyl group we can further impose $\phi_i \geq \phi_{i+1}$ and $\phi_{n-1} \geq |\phi_n|$. When reflections are taken into account, we are further allowed to set $\phi_n \geq 0$. Similarly, for $N_c=2n+1$
\begin{equation} \label{eq:eigenvalues phi N_c=2n+1}
\phi =  \mathrm{diag}(\phi_{1} , \ldots , \phi_{n} , -\phi_{1} , \ldots , -\phi_{n} , 0),
\end{equation}
with $\phi_i \geq \phi_{i+1} \geq 0$.

When $\phi_i$ are all different and non-vanishing the gauge symmetry is maximally broken to an Abelian group $U(1)^n$. For this reason the flat direction associated to vacuum expectation values (VEVs) of $\phi$ is called the Coulomb branch. By \eqref{eq:so(N_c) scalar potential}, whenever $\phi$ is as described above, all the matter fields are massive, except the bottom component of $Q$ when $N_c=2n+1$ (by \eqref{eq:eigenvalues phi N_c=2n+1}). Vector multiplets corresponding to off-diagonal generators are also massive (note that they are short multiplets satisfying a BPS bound). When some of the eigenvalues of $\phi$ are degenerate one can obtain an enhanced non-Abelian gauge symmetry.

There are also mixed branches where both $Q,\phi \neq 0$. This happens when some of the eigenvalues of $\phi$ vanish, leaving massless matter fields which are free to take expectation values that keep the first term of \eqref{eq:so(N_c) scalar potential} vanishing.

\subsection{Quantum aspects of the Coulomb branch}

\subsubsection{Fundamental instantons}

The classical moduli space described above typically does not remain intact in the quantum theory. The effective theory on the Coulomb branch is inflicted by non-perturbative dynamics, dominantly coming from instantons. In $d=3$ instantonic configurations are classified by $\pi_2$, which can be non-trivial on the Coulomb branch where the gauge group $G$ is typically broken to the maximal torus $U(1)^n$, leading to $\pi_2(G/U(1)^n) = \mathbb{Z}^n $. The topologically non-trivial configurations are precisely the static 't Hooft-Polyakov magnetic monopoles reduced to $d=3$, and accordingly carry magnetic charge under the unbroken $U(1)^n$.

When we consider the physics far away along the Coulomb branch and far from the points of enhanced gauge symmetry, the scale of the symmetry breaking (to the maximal torus) is high, the theory is weakly coupled and the instantonic dynamics can be described semi-classically. It was shown by Polyakov \cite{Polyakov:1976fu} that when all the topological sectors are taken into account the low energy $U(1)$ gauge theory is most naturally described by a scalar field. Let us recall how this scalar is related to the original gauge field.

In three dimensions, where a massless particle has no helicity, a free Abelian vector field can be dually described by a scalar. This statement is a lower dimensional version of the familiar electric-magnetic duality in $d=4$. Let us denote this dual scalar by $\gamma$. At the classical level
\begin{equation} \label{abelian duality}
F = - \frac{e^2}{2 \pi} * d \gamma,
\end{equation}
where $F$ is the field strength and $e$ is the Abelian gauge coupling. A special feature of three dimensional Abelian gauge theories is that the Bianchi identity can be interpreted as a statement of current conservation, since $*F$ has the correct quantum numbers to serve as a current. This symmetry usually goes by the name of $U(1)_J$. In the dual theory it acts by a shift of the dual scalar. In the quantum theory, charge quantization implies that the dual scalar takes values on a compact target space, that is by \eqref{abelian duality}
\begin{equation}
\frac{1}{e^2} \int * F  \in  \mathbb{Z}   \rightarrow    \gamma \in \frac{\mathbb{R}}{2\pi\mathbb{Z}}.
\label{eq:period of gamma}
\end{equation}

Let us now describe the semi-classical analysis of the theory on the Coulomb branch, starting from the purely bosonic case. Consider an $SO(3)$ gauge theory with a Higgs field $\phi$ in the vector representation. The low-energy Abelian theory has magnetic monopoles as explained above.  It was shown by Polyakov that when the magnetic interactions of the monopoles are taken into account in the dilute gas approximation, a sine-Gordon type potential is generated for the dual photon $\gamma$, and the theory is gapped. Note that the potential breaks the shift symmetry of the scalar.

The picture changes considerably when there are fermions in the story as in our case. A special role is played by the fermionic zero modes in the monopole background. Because of the zero modes, the path integral in a non-trivial topological sector can only be non-vanishing when it is dressed with insertions of fermionic fields. When the fermions transform under a global $U(1)$ symmetry group, the effective action will typically not be invariant under this symmetry due to the fermionic zero modes. Nevertheless, it turns out that a combination of this with the shift transformation of the dual scalar is a good symmetry of the effective action\footnote{This means that the global symmetry is actually broken spontaneously rather than explicitly, and the dual photon is the Goldstone boson.}. This fact is of considerable importance to the work presented here.

Consider adding now a complex fermion in the vector representation to the bosonic theory above with Yukawa interactions. It was shown in \cite{Affleck:1982as} that in this case the photon remains massless and an instanton induced interaction between the dual scalar and the fermions replaces the sine-Gordon potential. In the Prasad-Sommerfield limit, where the Higgs potential vanishes, the theory becomes $\mathcal{N}=2$ supersymmetric, and the instanton solution satisfies the BPS bound. In the supersymmetric theory, working with the dual photon means that instead of the vector multiplet we started with, we write an effective action for a chiral superfield $\Phi$, in which the dual photon and the neutral Higgs field combine to form a complex scalar $2\pi \phi / g^2 + i \gamma$. The instanton induced interactions then must come from a superpotential.

As mentioned above, due to fermionic zero modes, certain global transformations are only symmetries of the effective action when they are combined with the shift transformation of the dual photon. This means that the chiral superfield containing the dualized photon transforms under global symmetries in the effective theory. In this paper we will repeatedly use these symmetries and holomorphicity together with some consistency checks to determine the exact superpotential of the theory.

Let us give a classical example of how this works. Consider the pure $SO(3)$ gauge theory discussed above. In an instantonic background of one unit of magnetic charge the complex fermion has 2 zero modes. This means that every instantonic amplitude is as follows. The bosonic contribution is given by $\exp(-\frac{4\pi}{g^2}\phi + 2 i\gamma)$, where the first term in the exponent is the usual action in the Prasad-Sommerfield limit, and the second term appears because of the long range interactions of the monopoles. As explained above, since there are two fermionic zero modes, the amplitude can only be non-zero when there are two fermionic insertions, that is schematically the amplitude is
\begin{equation}
\lambda \, \lambda \, e^{-\frac{4\pi}{g^2}\phi + 2 i\gamma}.
\end{equation}
Now, $\lambda$ has R-charge $1$, but it is evident that this transformation by itself is not a symmetry of the effective action. Nor is the shift transformation of the dual photon $\gamma$. Nevertheless, a combination of these two transformations is a symmetry of the effective action. Explicitly, it is given by
\begin{equation}
\lambda \rightarrow e^{i \alpha} \lambda, \qquad \qquad \gamma \rightarrow \gamma - \alpha.
\end{equation}
By supersymmetry the superpotential must be a holomorphic function of the chiral superfield $\Phi$ defined above. This information is sufficient to determine the exact superpotential of the theory to be \cite{Affleck:1982as}
\begin{equation}
W = \exp[- 2 \Phi].
\label{SP SU(2)}
\end{equation}
Following \cite{Aharony:1997bx} it is convenient to define a new chiral superfield on which the global symmetries act linearly, which in the semi-classical limit obeys $Y_1 \simeq \exp(2 \Phi)$, such that $W=1/Y_1$. $Y_1$ is chosen to be large when $\phi$ is large. With this choice it has R-charge $(-2)$, and far away from the origin of the Coulomb branch the K\"ahler potential has small perturbative corrections, so to leading order at large $Y_1$, $K \approx \frac{g^2}{2 (2 \pi)^2} \Phi \Phi^\dagger = \frac{g^2}{2 (4 \pi)^2} \log(Y_1) \log(Y_1^\dagger)$.

Let's discuss now how this is generalized to the case where the unbroken gauge symmetry is $U(1)^n$. In this case there are $n$ so-called fundamental monopoles \cite{Weinberg:1979zt}, each one associated to a simple root. Every simple root is associated with a triplet containing $h_\alpha$ (the dual simple root) and two other elements in the algebra, $E_\alpha$ and $F_\alpha$, satisfying $[E_\alpha,F_\alpha]=h_\alpha$, $[h_\alpha,E_\alpha]=2E_\alpha$ and $[h_\alpha,F_\alpha]=-2F_\alpha$. This triplet is a copy of the $ \mathfrak{su}(2,\mathbb{C}) $ algebra in the algebra of our group, and allows us to embed the 't Hooft-Polyakov monopole into our theory.

The action of a fundamental monopole-instanton is given by
\begin{equation}
S = - \frac{1}{2}\mathrm{Tr}(h_\alpha \Phi) = -\frac{\pi}{ g^2} \mathrm{Tr}(h_\alpha\phi) - \frac{i}{2} \mathrm{Tr}(h_\alpha \gamma).
\end{equation}
We define as above chiral superfields corresponding to the fundamental monopoles
\begin{equation}
Y_\alpha \simeq \exp[\frac{1}{2} \mathrm{Tr}(h_\alpha \Phi)] = \exp [\frac{\pi}{g^2} \mathrm{Tr}(h_\alpha\phi) + \frac{i}{2} \mathrm{Tr}(h_\alpha \gamma)] .
\end{equation}
 The ${Y_\alpha}$ are a natural set of coordinates on the Coulomb branch at the quantum level \cite{Aharony:1997bx,deBoer:1997kr}.

The periodicity \eqref{eq:period of gamma} of $\gamma$ is now generalized to $\gamma \rightarrow \gamma + 4\pi w $, with $w$ an element of the weight lattice $\Lambda = \{w \in \mathfrak{t}$ such that $\mathrm{Tr}(h_\alpha w)\in \mathbb{Z}$ for all simple roots $\alpha \}$.
In the rest of this paper we will only be interested in matter in the $\mathbf{N_c}$ of $SO(N_c)$, and we will take the gauge group to be $O(N_c)$ (or $SO(N_c)$) rather than $Spin(N_c)$. This then means that we may only mod $\gamma$ by weights corresponding to single-valued representations of $SO(N_c)$.

We shall need the semi-classical relation of the fundamental monopole operators $Y_{\alpha}$ with the classical coordinates on the Coulomb branch \eqref{eq:eigenvalues phi N_c=2n} and \eqref{eq:eigenvalues phi N_c=2n+1}.
For $N_c=2n$ ($n \geq 3$) we have
\begin{equation}
Y_1 \simeq e^{\Phi_1 - \Phi_2},\qquad \ldots  \qquad Y_{n-1} \simeq e^{\Phi_{n-1} - \Phi_n}, \qquad Y_{n} \simeq e^{\Phi_{n-1} + \Phi_n}.
\label{mon oper. N_c=2n}
\end{equation}
For $N_c=2n+1$ ($n \geq 2$)
\begin{equation}
Y_1 \simeq e^{\Phi_1 - \Phi_2}, \qquad \ldots  \qquad Y_{n-1} \simeq e^{\Phi_{n-1} - \Phi_n}, \qquad Y_{n} \simeq e^{2 \Phi_n}.
\label{mon oper. N_c=2n+1}
\end{equation}
Recall that up to a normalization the real part of $\Phi_i$ is $\phi_i$ of equations \eqref{eq:eigenvalues phi N_c=2n} and \eqref{eq:eigenvalues phi N_c=2n+1}.

Next we must determine how $Y_\alpha$ is charged under the global symmetries. We have already noted that this information may be extracted from the counting of zero modes. We will also be considering matter fields so we need to compute the number of zero modes in an arbitrary representation for the fermions. This is given by the following formula (see \cite{Weinberg:1979zt,deBoer:1997kr})
\begin{equation}
N = \frac{1}{2} \sum_{w} \mathrm{sign}(\mathrm{Tr}(w \phi))\mathrm{Tr}(w g),
\end{equation}
where $g$ is the magnetic charge of the monopole background (for the fundamental monopoles it is $h_\alpha$), and $w$ are all the weights in the representation.

We consider either ``gauginos'' in the adjoint or ``quarks'' in the fundamental.
The results for the $O(N_c)$ ${\cal N}=2$ theory with $N_f$ ``quarks'' may be summarized as follows. The gauginos have two zero modes for each fundamental monopole. For the quarks the result depends on the structure of the algebra.
For $N_c=3$ the quarks are the same as the gauginos.
For $N_c=2n$ ($n \geq 2$), if $\phi_n > 0$ ($\phi_n < 0$) each quark has 2 zero modes for $Y_n$ ($Y_{n-1}$), and none for the other monopoles, implying global charges
\begin{longtable}{|c|c|c|}
\hline
 & $U(1)_{R}$  & $U(1)_{A}$\tabularnewline
\hline
\hline
$Y_i$  & $-2$  & $0$\tabularnewline
\hline
$Y_n (Y_{n-1})$  & $2N_f-2$  & $-2N_f$\tabularnewline
\hline
\end{longtable}
For $N_c=2n+1$ ($n \geq 2$) each quark has 2 zero modes for $Y_n$. Hence
\begin{longtable}{|c|c|c|}
\hline
 & $U(1)_{R}$  & $U(1)_{A}$\tabularnewline
\hline
\hline
$Y_i$  & $-2$  & $0$\tabularnewline
\hline
$Y_n $  & $2N_f-2$  & $-2N_f$\tabularnewline
\hline
\end{longtable}

There is another instantonic operator, that naturally appears if we consider our $d=3$ theory as the limit
of a $d=4$ SQCD theory compactified on a small circle.
In this case one may ask for the full classification of finite action configurations, that contribute in the semi-classical approximation.
We have already discussed at length the configurations associated with $d=3$ monopoles which are independent of the compactified dimension. This does not exhaust the set of all semi-classical contributions \cite{Gross:1980br}. An additional configuration has negative magnetic charge given by $(- \sum_\alpha k_\alpha h_\alpha)$, where $k_\alpha$ are such that $\sum_\alpha k_\alpha h_\alpha$ is the shortest dual root. For the cases which are of interest to us, they are given by $(1,2,\ldots,2,1,1)$ for $N_c=2n$ ($n \geq 3$) and $(1,2,\ldots,2,1)$ for $N_c=2n+1$ ($n \geq 2$). Together with the fundamental instantons, these give the full set of fundamental semi-classical configurations for the $d=4$ theory on a circle. In fact the instanton of the $d=4$ gauge theory is found to be a bound state of these two types of configurations \cite{Lee:1997vp,Lee:1998vu}.

This motivates the definition in the $d=3$ theory of an instantonic operator which has the aforementioned magnetic charge
\begin{equation}
Z \simeq \exp[\sum_\alpha k_\alpha \mathrm{Tr}(h_\alpha \Phi)].
\label{twist mon}
\end{equation}
For $N_c=2n$ this is
\begin{equation}
Z = Y_1 Y_2^2 \times \ldots \times Y_{n-2}^2 Y_{n-1} Y_n,
\label{Z def N_c=2n}
\end{equation}
and for $N_c=2n+1$
\begin{equation}
Z = Y_1 Y_2^2 \times \ldots \times Y_{n-1}^2 Y_n.
\label{Z def N_c=2n+1}
\end{equation}
In view of \eqref{mon oper. N_c=2n} and \eqref{mon oper. N_c=2n+1} we have that semi-classically $Z\simeq e^{\Phi_1 + \Phi_2}$ for all values of $N_c$.

When the $d=4$ theory is placed on a circle the leading correction to the superpotential associated with the circle comes from these semi-classical configurations \cite{Seiberg:1996nz,Aharony:1997bx,Davies:1999uw,Davies:2000nw}, and is given by $\eta_{N_c,N_f} Z$, with $\eta_{N_c,N_f} = \Lambda_{N_c,N_f}^{3(N_c-2)-N_f}$ ($N_c>4$, where $\Lambda_{N_c,N_f}$ is the strong coupling scale of the $d=4$ gauge theory). From the relation of $Z$ to the $Y_\alpha$ given above we can find that $Z$ has R-charge $(2N_f-2N_c+6)$ and axial charge $(-2N_f)$ for all values of $N_c$. In the $d=4$ theory on a circle, the superpotential $\eta_{N_c,N_f} Z$ breaks these symmetries to a linear combination $\tilde{\mathrm{R}}$ such that $Q$ has charge $\frac{N_f-N_c+2}{N_f}$. This is precisely the correct non-anomalous symmetry transformation in $d=4$.

In the $U(N_c)$ and $USp(2N_c)$ gauge theories, it was found in \cite{Aharony:1997bx,Karch:1997ux,Aharony:1997gp} that for all values of $N_f$ most
of the Coulomb branch is lifted by quantum corrections, that at most a one-dimensional subspace of
the Coulomb branch remains, and that this one-dimensional subspace can be naturally parameterized by the chiral superfield $Z$. In those cases $Z$ is globally well-defined and seems to be a good variable in
the quantum theory near the origin of the moduli space. In the $O(N_c)$ theories we discuss in this paper,
we will see that $Z$ does not provide a good variable. This is related to the fact that in these theories
the lattice of allowed magnetic charges for local operators is not spanned by the 't Hooft-Polyakov monopole operators $Y_{\alpha}$ described above, but there is a smaller monopole charge that is allowed by the Dirac quantization condition \cite{Kapustin:2005py} (the quantization condition is reviewed in appendix \ref{appB})\footnote{We thank C. Closset and S. Cremonesi for discussions on this issue.}. For low-rank gauge groups this can easily be understood. In the $O(3)$ theory the allowed electric charges are only integers, when we normalize charge so that the $SU(2)$ theory has also half-integer charges, so a magnetic charge that is half that of the $SU(2)$ 't~Hooft-Polyakov monopole is also allowed. In the $O(4) \simeq (SU(2)\times SU(2))/\mathbb{Z}_2$ theory,
half-integer electric charges are allowed in the two $SU(2)$'s, but only when both electric charges are non-integer, allowing a magnetic configuration which carries half of the minimal $SU(2)$ magnetic charge under both gauge groups (but not under a single gauge group). One can show that in all cases the quantum numbers of the operator with minimal magnetic charge are those of $\sqrt{Y_1 Z} \simeq e^{\Phi_1}$. There is no semi-classical 't Hooft-Polyakov monopole that carries this charge, and apriori it is not clear how
many fermionic zero modes this minimal monopole operator carries. We will conjecture that there is a chiral superfield $Y$ (with $Y^2 = Y_1 Z$) in the $O(N_c)$ theories that carries this minimal magnetic charge, and we will argue that this superfield is the correct variable to use in the quantum theory to parameterize what remains of the Coulomb branch (in a manner that is non-singular near the origin of the moduli space).

\subsubsection{Chern-Simons term parity anomaly and real masses}

In the last section of this paper we will also discuss dualities of Chern-Simons gauge theories.
In ${\cal N}=2$ the Chern-Simons term has the following supersymmetric completion :
\begin{equation}
\label{lcs}
\mathcal{L}_{\mathrm{CS}} = \frac{k}{4 \pi} \mathrm{Tr} \left( A \wedge \mathrm{d}A + \frac{2i}{3} A \wedge A \wedge A - \lambda \bar{\lambda} + 2 \phi D \right).
\end{equation}
Under gauge transformations not homotopic to the identity, with winding number $n$, the Chern-Simons action is shifted by $2 \pi k n$. Hence, for $e^{iS_{\mathrm{CS}}}$ to be gauge-invariant we must have that $k \in \mathbb{Z}$.

Consider an $SU(N_c)$ gauge theory with one fermion in the fundamental representation. It was shown in \cite{Niemi:1983rq,Redlich:1983dv}, that under a homotopically non-trivial gauge transformation, the fermion determinant changes according to $\det[\gamma^{\mu} \mathcal{D}_\mu] \rightarrow  (-1)^{|n|} \det[\gamma^{\mu} \mathcal{D}_\mu]$. This implies that the effective action, for an odd number of fermions, is gauge-invariant only when a Chern-Simons term with $k \in \mathbb{Z}+\frac{1}{2}$ is included to compensate the shift to the determinant.

The parity anomaly gives a $\mathbb{Z}_2$ version of 't Hooft anomaly matching in $d=4$ \cite{Aharony:1997bx}. If we weakly gauge a global symmetry group then, depending on the content of charged matter, for the quantum theory to be consistent, we either have to include a Chern-Simons term with an half-integer coefficient or we do not. This information must match for dual theories (which have the same global symmetries). As an example, for a $U(1)_R \times U(1)_A$ anomaly, each fermion gives $\frac{1}{2} (\mathrm{R \, charge}) \times (\mathrm{A \, charge})$, and the sum of all contributions can be either an integer or an half-integer.

For the gauge theories we will be discussing in this paper, $SO(N_c)$ and $O(N_c)$, there is no parity anomaly. This can be seen by considering the $SO(3)$ case. For a fermion in the $\mathbf{2}$ representation the fermion determinant transforms by a minus sign as discussed above, but for a fermion in the adjoint, the vector of $SO(3)$, the determinant does not change sign. This means that there is no sign ambiguity for the determinant and no parity anomaly. This result holds for all values of $N_c$, and implies that in $SO(N_c)$ or $O(N_c)$ theories the number of flavors $N_f$ does not have to be even.

There is a deep connection between real masses for fermions and the Chern-Simons term. Integrating out a fermion $\psi$, in the $\mathbf{N_c}$ of $SO(N_c)$ or $O(N_c)$ with a real mass $m_{\psi}$, shifts the Chern-Simons level, by a one-loop effect, according to $k \rightarrow k + \mathrm{sign}[m_{\psi}]$ \footnote{For fermions in the fundamental of $SU(N_c)$ the shift is $k \rightarrow k + \frac{1}{2}\mathrm{sign}[m_{\psi}]$, consistent with the parity anomaly. For $SO(N_c)$ and $O(N_c)$ theories, for which we normalize ${\rm tr}(T^a T^b) = 2 \delta^{ab}$, we take the coefficient in \eqref{lcs} to be $k/16\pi$. For $N_c=3$ this normalization is not the natural one, but using the same normalization for all values of $N_c$ simplifies our discussion of flows which change $N_c$, and this issue will not play any role in this paper.}.  This allows us to relate theories with different Chern-Simons levels by a renormalization group flow induced by a real mass deformation, as discussed in detail in \cite{ItamarThesis}.

\subsubsection{The pure super-Yang-Mills theory}

We shall first consider the case with $N_f=0$, that is the pure supersymmetric Yang-Mills (SYM) theory. At the classical level there is a Coulomb branch given by  \eqref{eq:eigenvalues phi N_c=2n} and \eqref{eq:eigenvalues phi N_c=2n+1} and the dual photons. Quantum mechanically, we ask whether a superpotential can be generated by non-perturbative effects, as in the $SU(2)\simeq SO(3)$ case \cite{Affleck:1982as}. The relevant degrees of freedom to consider are the $n=\mathrm{rank}(O(N_c))$ fundamental instantons. By the zero mode counting described in the previous section, we know the R-charge of all these operators. When $N_f=0$ they all have R-charge $(-2)$. Demanding that the superpotential has R-charge $2$, we write the following superpotential
\begin{equation}
W = \sum_{i=1}^{n} \frac{2}{\mathrm{Tr}(\alpha_i ^2)} \frac{1}{Y_i}.
\label{SP YM}
\end{equation}
This superpotential was shown to be the correct one by explicit computation in \cite{Davies:2000nw}, from which we have adopted the normalization up to global factors\footnote{Note that $\mathrm{Tr}(\alpha_i ^2)=2$ for all the simple roots, except for $\alpha_n$ of $B_n$ ($N_c=2n+1$) for which $\mathrm{Tr}(\alpha_n ^2)=1$. This is important for getting the correct value of the $d=4$ gaugino condensation as we explain below.}. Similarly to the pure $SO(3)$ case, this potential implies a runaway behaviour for each of the classical flat directions. Thus, this theory does not have a stable vacuum.

A nice check of this superpotential is to consider the $d=4$ pure SYM theory compactified on a circle. We have already mentioned that the leading correction to the superpotential is proportional to the $Z$ operator. We have then
\begin{equation}
W_{S^1\times \mathbb{R}^3} = \sum_{i=1}^{n} \frac{2}{\mathrm{Tr}(\alpha_i ^2)} \frac{1}{Y_i}+ \eta_{N_c,0} Z,
\end{equation}
with $\eta_{N_c,0}=\Lambda^{3(N_c-2)}$. Solving $\partial W / \partial Y_i = 0$ to find vacuum configurations, and taking into account the difference between \eqref{Z def N_c=2n} and \eqref{Z def N_c=2n+1}, we find for both cases $(2 \eta Z)^{N_c-2}=2^4 \eta$. This implies that there are exactly $(N_c-2)$ vacua in $d=4$, as expected \cite{Witten:1982df,Intriligator:1995id,Witten:1997bs}. Moreover, plugging the solution back into the superpotential gives $W=\frac{N_c-2}{2} \epsilon_{N_c-2} 2^{\frac{4}{N_c-2}} \Lambda^3$, which gives the correct value of the $d=4$ gaugino condensate ($\epsilon_{N_c-2}$ are the $N_c-2$ roots of unity).

For $N_f > 0$, as in the $U(N_c)$ and $USp(2N_c)$ cases, the superpotential \eqref{SP YM} is still generated for all but one of the fundamental instantons which has more zero modes, lifting the Coulomb branch except possibly for a one-dimensional subspace. Apriori it is not clear which coordinate should be used to parameterize this
one-dimensional subspace -- $Z$, $Y$ or some other superfield (of course this makes no difference far out along the Coulomb branch, but it makes a difference near the origin of the Coulomb branch where some choices of variables have a singular K\"ahler metric while others do not). We will start in the next
section by analyzing the dynamics for large values of $N_f$, where we will see that a dual description exists if we use the $Y$ variable, but not for other choices. We thus claim that $Y$ is the correct
variable to use to parameterize the Coulomb branch in these theories, for all values of $N_f$ (since one
can flow to lower values from the large $N_f$ theories). It would be interesting to find a direct justification for this.

\section{An $O(N_f-N_c+2)$ dual theory}
\label{duality}

In this section we will discuss a proposal for a dual description for the $O(N_c)$ SQCD theories
with $N_f \geq N_c$ \footnote{As mentioned in the introduction, the same duality was also recently suggested in \cite{Benini:2011mf,Hwang:2011ht}.}. For these values of $N_f$ the gauge group is completely broken at generic
points on the Higgs branch. As mentioned above most of the Coulomb branch is lifted by instanton-generated superpotentials (as for all other $d=3$ ${\cal N}=2$ gauge theories), and at most a one-dimensional
subspace remains. By looking at the quantum numbers of the various possible Coulomb branch coordinates, it is easy to verify that no superpotential can lift this part of the Coulomb branch, so the moduli space of the quantum theory consists of the Higgs branch and of a one-dimensional Coulomb branch, with a superconformal fixed point at the origin of moduli space.
% where the two branches intersect.

Based on the brane construction \cite{Elitzur:1997fh} it is natural to guess that a dual theory (flowing to the same
fixed point) should involve an $O(N_f-N_c+2)$ gauge group (noting the charges of the orientifolds
involved), and that its form should be similar to that of other $d=3$ ${\cal N}=2$ dualities, namely
it should include as singlet fields the mesons and some Coulomb branch coordinates of the original
theory, with a superpotential coupling them to the mesons and Coulomb branch coordinates of the dual theory. The most non-trivial element in constructing such a duality is the identification of the proper gauge-invariant degrees of freedom that are needed for the description of the Coulomb branch.
In the previous section we discussed the description of the Coulomb branch at the quantum level. When the corrections due to instantons are taken into account, it is found that the natural coordinates for describing the effective theory on the Coulomb branch are the monopole operators. In the previous section we discussed two natural choices of coordinates on the Coulomb branch, $Z$ (which naturally appears in the $d=4$ theory on a circle) and $Y$ (which is the monopole operator with the minimal magnetic charge).
Apriori it is not clear which one should behave nicely near the origin of moduli space, so that it should be a good variable to use in the effective description. However, our attempts to construct a sensible dual/effective description using $Z$ did not work, so we will advocate in this section the operator $Y$ as
the correct degree of freedom for describing the residual Coulomb branch. We will exhibit several checks of this proposal.

Let us discuss in detail the proposed duality for the $O(N_c)$ gauge theory. We suggest that, for $N_f \geq N_c$, it is dual to an $O(\tilde{N_c})$ gauge theory ($\tilde{N_c}=N_f-N_c+2$), with $N_f$ chiral superfields $q$ in the $(\mathbf{N_f-N_c+2})$ representation. In the IR, the unlifted Coulomb branch of the dual theory is parameterized by $\tilde{Y}$. In addition, two gauge-invariant chiral superfields are part of the dual theory: $M$ and $Y$. The former is identified with the mesons of the original theory, while the latter is identified with the $Y$ monopole operator parameterizing the Coulomb branch of the original theory. This leads to the following quantum numbers under the global symmetries :

\begin{table}[h]
\begin{center}
\begin{tabular}{c|c c c}

 & $U(1)_{R}$  & $U(1)_{A}$ & $SU(\mathbf{N_f})$\tabularnewline
\hline
\tabularnewline
$Y$  & $N_f-N_c+2$  & $-N_f$ & $\mathbf{1}$\tabularnewline\tabularnewline

$M$ & $0$ & $2$ & $\mathbf{\frac{N_f(N_f+1)}{2}}$\tabularnewline\tabularnewline

$\tilde{Y}$  & $N_c-N_f$  & $N_f$ & $\mathbf{1}$\tabularnewline\tabularnewline

$q$ & $1$ & $-1$ & $\overline{\mathbf{N_f}}$\tabularnewline

\end{tabular}
\end{center}
\end{table}
We chose the $U(1)$ charges and $SU(N_f)$ representation of $q$ such that a superpotential
of the form
\begin{equation}
W = \frac{1}{2} q_i M^{ij} q_j + Y\tilde{Y},
\label{generalW}
\end{equation}
which is similar to the one suggested in \cite{Aharony:1997gp}
for the symplectic gauge group duality, is consistent with all the global symmetries. Note that the charges of
${\tilde Y}$ are fixed given the charges of $q$.

Because we are considering theories with $O(N_c)$ gauge group, where reflections are gauged, the baryons are not gauge-invariant operators, and hence do not play any role in the duality. For the $SO(N_c)$ theory, including the baryon operators, we could not in general find any suggestion for a dual description. Nevertheless, we will see below that, at least in one simple case, the duality can be generalized to the special orthogonal group $SO(N_c)$. In this case the baryons will turn out to be involved in a non-trivial way.

We can perform several tests of the duality. First, performing the duality transformation twice brings us back to the same theory we started from. Next,
the duality is consistent with complex mass deformations and Higgsing, in a way similar to other examples of Seiberg duality. Adding a complex mass term $W=\frac{1}{2}{\rm tr}(mM)$ with $\mathrm{Rank}(m)=l$ reduces the number of massless flavors in the original theory to $N_f-l$. In the dual theory, the equations of motion for the massive mesons imply the dual gauge group is broken with $\langle q_i q_j  \rangle=-m_{ij}$, and therefore only an $O(N_f-N_c-l+2)$ gauge group remains with $N_f-l$ massless flavors, consistent with the duality.

Similarly, going on the Higgs branch of the original theory, with non-vanishing expectation values for $M$, corresponds to giving complex masses to the dual quarks. In the original theory we must have that $\mathrm{Rank}(M) \leq N_c$, but in the dual theory, naively, $\mathrm{Rank}(M)$ is not constrained. The resolution of this difficulty, which is standard in Seiberg dualities, is that for VEVs of higher rank a runaway superpotential is generated, so they do not lead to supersymmetric vacua in the dual theory.
In our case, when $\mathrm{Rank}(M)$ exceeds $N_c$ we obtain an $O(N_f-N_c+2)$ gauge theory with $\tilde{N_f}= N_f-\mathrm{Rank}(M)$ massless flavors, such that $\tilde{N_f}< \tilde{N_c}-2$. We will see below
that indeed whenever
$N_f < N_c-2$ a runaway superpotential is generated and there are no supersymmetric vacua.
Thus, noting that the superpotential lifts the standard Higgs and Coulomb branches of the dual theory, the two theories have identical moduli spaces. Related to this, the list of chiral gauge-invariant operators in the two theories is the same (containing $M$, $Y$ and their products).

A less obvious test of the dualities is given by the parity anomaly matching reviewed in section 2. We only know how to compare the theories at the UV, so
we will compare the anomalies for $W_{\alpha}$ and $Q$ in the original theory with those of $\tilde{W}_{\alpha}$, $q$, $M$ and $Y$ in the dual theory. The comparison is as follows (OT = original theory, DT = dual theory) :
\begin{align}
& U(1)_{R}^{2}   & OT = & \frac{1}{2} N_cN_f+\frac{1}{4}N_c(N_c-1) \nonumber \\
& & DT = & \frac{(N_f-N_c+2)(N_f-N_c+1)}{4}+\frac{N_f(N_f+1)}{4}+\frac{(N_f-N_c+1)^{2}}{2}\\
%\end{align}
%\begin{align}
& U(1)_{A}^{2}   & OT = & \frac{1}{2}N_cN_f \nonumber\\
& & DT = & N_f(N_f+1)+\frac{1}{2}(N_f)^{2}+\frac{1}{2}(N_f-N_c+2)N_f\\
%\end{align}
%\begin{align}
& U(1)_{R} \times U(1)_{A}  & OT = & -\frac{1}{2} N_cN_f \nonumber\\
& & DT = & -\frac{1}{2}N_f(N_f+1)+\frac{1}{2}(N_f-N_c+1)(-N_f)
\end{align}
\begin{align}
& SU(N_f)^2  & OT = & \frac{1}{2} N_c \nonumber\\
& & DT = & \frac{1}{2}(N_f-N_c+2)+\frac{1}{2}(N_f+2).
\end{align}
Consistency requires that these should match modulo one, as
they do indeed.

The duality above is also consistent with the duality for theories with a Chern-Simons term of level $k$, suggested in \cite{Kapustin:2011gh}, which maps $O(N_c)_k$ gauge groups to $O(N_f-N_c+2+k)_{-k}$. As mentioned above, we can flow to a non-zero value of $k$ by adding real masses, and we will analyze this in detail in section 5.

\subsection{$N_f=N_c=2$ -- an Abelian-Abelian duality}

The case of $N_f=N_c$ is of particular interest since the dual gauge group is an Abelian $O(2)$ group,
which is closely related to the $U(1)$ gauge theory for which a different duality (to a $U(N_f-1)$ theory) has been proposed in \cite{Aharony:1997gp} (the $O(2)$ theory is the same as a $U(1)$ gauge theory with
the charge conjugation symmetry gauged). For an Abelian $SO(2)$ (or $U(1)$) gauge theory, the flavor symmetry group is enhanced to $SU(N_f)\times SU(N_f)\times U(1)_A$, with a separate flavor symmetry acting on $Q_1^i+iQ_2^i$ ($i=1,\cdots,N_f$) and on $Q_1^i-iQ_2^i$, and there is also a topological $U(1)_{J}$ symmetry. The duality we proposed above does not include these extra symmetries (though they might arise as accidental symmetries on one side of the duality). This is resolved by going to $O(N_c)$ theories, since the $O(2)$ theory has an extra gauge transformation which is charge conjugation, and only an $SU(N_f)\times U(1)_A$ global symmetry commutes with this extra gauge transformation. This gives another reason for why we need the
general orthogonal group $O(N_c)$ (in contrast to the special one $SO(N_c)$) for the duality to work.

In particular it is interesting to discuss the $N_f=N_c=2$ case which is an Abelian-Abelian duality. We will analyze this case both for $SO(2)$ and for $O(2)$ gauge groups, and we will see that in this specific case the duality works also for $SO(2)$, and that our
 duality is closely related to one of the $U(N_c)$ dualities considered in \cite{Aharony:1997gp}, between two $U(1)$ $N_f=2$ theories.

Recall how the general $U(N_c)$ duality works. In the electric theory there are matter fields $P$ in the $\mathbf{N_c}$ representation of $U(N_c)$ and $\tilde{P}$ in the $\overline{\mathbf{N_c}}$ representation, and mesons defined by $M_P=P \tilde{P}$, which are in the bi-fundamental representation of the $SU(N_f)\times SU(N_f)$ flavor group. Additionally, there are monopole operators $V_{\pm}$ parameterizing the residual Coulomb branch (which is split into two by its intersection with the Higgs branch \cite{Aharony:1997bx}). The dual theory has a gauge group $U(N_f-N_c)$ and matter fields $p$ and $\tilde{p}$ in the $\mathbf{N_f-N_c}$ and $\overline{\mathbf{N_f-N_c}}$ representations, respectively. There are singlet fields $M_P$ and $V_{\pm}$, monopole operators $\tilde{V}_{\pm}$ of the dual gauge group, and a superpotential
\begin{equation}
W = p M_P \tilde{p} + V_{+} \tilde{V}_{-} + V_{-} \tilde{V}_{+}.
\label{U(N_c) duality}
\end{equation}

We now focus on the duality for $U(1)$ with $N_f=2$, and use the $SO(2)$ language to describe it. Instead of $P$ and $\tilde{P}$ we have $Q$ in the $\mathbf{2}$ of $SO(2)$. The $N_f\times N_f$ matrix $M_P$ decomposes into a symmetric part $M^{ij}=Q_a^i Q_a^j$ and an antisymmetric part $B^{ij} = \epsilon^{ab}Q_a^i Q_b^j$, which is a baryon operator from the point of view of the $SO(2)$ theory. Similarly, in the dual theory there are $\tilde{B}^{ij}=\epsilon^{ab}q_a^i q_b^j$, such that $p M_P \tilde{p} = q M q - B \tilde{B}$.
Thus, the $U(1)$ duality gives us a dual $SO(2)$ theory with singlet fields $V_{\pm}$, $M$ and $B$, and with a superpotential
\begin{equation}
W = V_{+} \tilde{V}_{-} + V_{-} \tilde{V}_{+} + q M q - B \tilde{B}.
\label{specialW}
\end{equation}

So, in this special case we have a dual description also for $SO(2)$, when the baryons are properly taken into account. We can now obtain a duality for $O(2)$
by imposing an identification by the charge conjugation symmetry, and we claim that this gives precisely the $O(N)$ duality that we described above. Indeed, this symmetry takes
$B\to -B$, $\tilde{B}\to -\tilde{B}$, $V_{\pm} \to V_{\mp}$ and ${\tilde V}_{\pm} \to {\tilde V}_{\mp}$; after imposing this, $Y = V_+ = V_-$
and ${\tilde Y} = {\tilde V}_+ = {\tilde V}_-$ remain as the only gauge-invariant Coulomb branch coordinates, and the superpotential
\eqref{specialW} becomes exactly the same as our general superpotential \eqref{generalW}.

When we consider $N_f=N_c$ with $N_c>2$, we still have a dual $O(2)$ theory, but there is no similar
duality for the $SO(N_c)$ theory.
Similarly, our dualities which take an $O(2)$ theory to an $O(N_f)$ theory do not work for the gauge group $SO(2)$. Nevertheless,
the $O(2)$ theory can still be obtained from an $SO(2)=U(1)$ theory by gauging charge conjugation, and we
can identify this symmetry on both sides of the $U(N_c)$ duality, implying that there is still a relation between
the two dual theories (the one from $U(N_c)$ duality and the one from $O(N_c)$ duality) in these cases.

\subsection{$N_f=N_c-1$ duality} \label{N_f=N_c-1}

In this subsection we provide a dual description for $N_f = N_c-1$. A naive application of the
general duality above would give us a dual $O(1)$ theory in this case, but since this symmetry is discrete, there is no Coulomb branch so we do not have a variable $\tilde{Y}$ that we can use. We will see that there is indeed an $O(1)$ dual theory, but some modification of the general duality will be required.
The most important clue is that as we go along the Higgs branch with $\mathrm{Rank}(M)=N_c-2=N_f-1$ we flow at low energies to an $O(2)$ theory with $N_f=1$, and we should recover the known description of this theory (as an identification of the $U(1)$ theory with $N_f=1$ by charge conjugation). We will see that the dual chiral superfields $q$, now charged under the discrete dual gauge group $O(1)$, are crucial for this matching.

Recall that for $SO(2) \cong U(1)$ gauge theories with flavors the Coulomb branch, which is classically a cylinder, is split by quantum effects into two half-cylinders, intersecting each other and the Higgs branch at the origin of moduli space \cite{Aharony:1997bx,deBoer:1997kr}. Each half of the Coulomb branch may be parameterized by a separate variable $V_{\pm}$, and these variables carry charge $\pm 1$ under the topological $U(1)_J$ symmetry.
In the special case of $N_f=1$, the Higgs branch has one complex dimension parameterized by $M$, and
the quantum picture of the moduli space is topologically that of three half-cylinders $V_{\pm},M$ intersecting at a point, where there is a conformal fixed point.

Another theory which flows to the same interacting fixed point can be given in terms of the three chiral superfields $V_{\pm},M$ and a superpotential $W = M V_{+} V_{-}$ \cite{Aharony:1997bx}. This superpotential correctly captures the moduli space of three intersecting cones. The quantum numbers are

\begin{longtable}{c|ccc}

 & $U(1)_{R}$  & $U(1)_{A}$ & $U(1)_{J}$\tabularnewline
\hline

$M$ & $0$ & $2$ & $0$\tabularnewline

$V_{\pm}$  & $1$  & $-1$ & $\pm1$\tabularnewline

\end{longtable}

Let us now try to find a low-energy effective description
of the moduli space for the general case of $N_f=N_c-1$, both for $SO(N_c)$ and for $O(N_c)$ gauge groups. Since at a generic point in the moduli space the gauge group is completely broken, one may guess that the chiral fields $M$ and $Y$ are sufficient for a description of the low-energy effective theory. It is amusing to pursue this possibility, before discovering that new degrees of freedom must become massless at the origin for the description to be consistent with the flow to the $SO(2)$ theory with $N_f=1$, and with
the mass flow from the theories with higher $N_f$ considered above.

For $N_f=N_c-1$, $Y$ has R-charge $1$ and axial charge $(-N_f)$. There is then (unlike the case for higher values of $N_f$) a unique smooth superpotential that one can write, consistent with the global symmetries,
\begin{equation}
W=Y_{N_c,N_c-1}^{2}\det(M).
\end{equation}
However, if we assume an effective description using this superpotential, and take
large VEVs for $(N_f-1)$ of the eigenvalues of $M$, we do not recover the results we expect for $SO(2)$ or $O(2)$ with one flavor (assuming that $Y$ and $M$ are identified in some proper way with $V_{\pm}$; in the $O(2)$ case we expect $V_+$ to be identified with $V_-$, but two half-cylinders should still remain in the moduli space).
For simplicity, let's show this for the $N_c=3$ case. We may choose
to focus on the part of the Higgs branch where $M$ is diagonal
with $M_{22}\gg M_{11}$. The F-term equations are
\begin{equation}
2Y_{3,2}M_{11}M_{22} =  Y_{3,2}^{2}M_{11} = Y_{3,2}^{2}M_{22}=0.
\end{equation}
The $Y$ coordinate of the low-energy theory is given by the standard matching $Y_{2,1}^2 = Y_{3,2}^2 M_{22}$.
Obviously, for large $M_{22}$ there is no branch where $Y_{2,1}\neq0$, while as described above we do expect a Coulomb branch to remain in this case.
One can also check that $Y$ and $M$ by themselves are not sufficient to match the parity anomalies.

The solution to this is to introduce the dual chiral superfields $q_i$ ($i=1,\cdots,N_f$) as above, and to write down the most general superpotential consistent with the global symmetries,
\begin{equation}
W=Y_{N_c,N_c-1}^{2}\det (M)+ \frac{1}{2} q_i M^{ij} q_j.
\label{effdesc}
\end{equation}
The second term arises when we flow from the theories with higher values of $N_f$, and we conjecture that the first term is also generated by quantum effects, since it is allowed by the symmetries.
We already know from the general case that the anomaly matching conditions are now satisfied; note that the fundamental fields used for the anomaly matching are the same here as in the general case, despite the fact that we have no ${\tilde Y}$ field.

We can also verify that this theory indeed flows in the correct way to the $O(2)$ theory with $N_f=1$.
When $M$ takes on an expectation value of rank $N_f-1$, $q_{i}$ ($i=2,\cdots,N_f$) receive a mass, and after they are integrated out we are left with the superpotential
\begin{equation}
W=M^{11}(Y_{2,1}^{2}+ \frac{1}{2} q_{1}^{2}),
\label{flow to so(2)}
\end{equation}
where we used the matching condition $Y_{2,1}^{2}=Y_{N_c,N_c-1}^{2}\mathrm{subdet}(M)$.
In the $SO(N_c)$ case the resulting theory has an accidental $U(1)_J$ symmetry rotating $Y_{2,1}$
and $q_{1}$, and it is natural to define
\begin{equation}
V_{\pm}=Y_{2,1}\pm \frac{1}{\sqrt{2}} iq_{1}.
\label{vident}
\end{equation}
We can now identify
\eqref{flow to so(2)} with $W=M^{11} V_+ V_-$, and obtain precisely the expected low-energy description. Note that under this assignment $V_{\pm}$ has the expected R-symmetry and axial charges. The effective description
\eqref{effdesc} thus seems to work also for the $SO(N_c)$ case (note that since $N_f < N_c$ there are no
baryon operators in this case).

In the $O(N_c)$ theory we must still impose that reflections are gauged. For the original theory this implies that on the moduli space $V_+$ is identified with $V_-$. For the dual theory the gauge group is $O(1)$, which acts by taking $q$ to $-q$. These identifications are consistent with \eqref{vident}, providing a consistency check for our effective description.

Note that for higher values of $N_f$ there are still subspaces of the Higgs branch with an unbroken $O(2)$ symmetry, and on these subspaces the effective theory should be closely related to the low-energy limit of a $U(1)$ gauge theory. As mentioned in the previous subsection, this has an alternative description using the $U(N_c)$ duality of \cite{Aharony:1997gp}, and the descriptions should be consistent with each other using the relation between the $U(N_c)$ and $O(N_c)$ dualities described in the previous subsection.

\section{Theories with $N_f \leq N_c -2$}

In this section we briefly discuss how to flow to theories with a lower number of flavors.
 Let us first consider the case of $N_f = N_c - 2$. In this case, generically on the Higgs branch we obtain an $O(2)$ theory with no flavors, so we expect the Coulomb branch to be unlifted and the quantum moduli space to be parameterized by $Y$ and $M$. Since there is no dual gauge group in this case, we expect that $M$ and $Y$ should be sufficient to obtain a low-energy description. In particular, in this case both $M$ and $Y$ have no R-charge, implying that no superpotential can arise on the moduli space.

 Nevertheless, the quantum moduli space does not have to be the same as the classical moduli space; as in similar cases in which the generic low-energy theory is a free $U(1)$ gauge theory in $d=4$ and in $d=3$
 \cite{Seiberg:1994rs,Seiberg:1996nz}, we expect that the quantum moduli space may be modified so that the point at the origin $Y=M=0$ is no longer singular. Note that the global charges of $Y$, which has R-charge 0 and axial charge $(-N_f)$, allow us to write a neutral combination $Y^2 \det(M)$.

To see how the moduli space is deformed, we can start
with the $N_f = N_c - 1$ theory and integrate out one flavor. Adding a mass term we find that
the F-term equation of $M^{N_f N_f}$ takes the form (in terms of the meson matrix for the
remaining $N_c-2$ massless flavors)
\begin{equation}
Y_{N_c,N_c-1}^2 \det(M) + \frac{1}{2} q_{N_f}^2 + \frac{1}{2}m = 0.
\label{N=F-2}
\end{equation}
Using the natural rescaled Coulomb branch coordinate $Y_{N_c,N_c-2}^2 = 2 Y_{N_c,N_c-1}^2 / m$, and
defining ${\tilde q} = q_{N_f} / \sqrt{m}$, we can rewrite this as
\begin{equation}
Y_{N_c,N_c-2}^2 \det(M) + {\tilde q}^2 + 1 = 0.
\label{constraint}
\end{equation}
Generically ${\tilde q}$ is non-zero, so that all other variables (the $M^{N_f i}$ and the other $q_i$'s) are massive, and we remain only with the fields appearing in \eqref{constraint}, which are constrained by this equation. This agrees with our general expectation that the moduli space is not lifted, since any value for $Y$ and $M$ is allowed (and $\tilde q$ is not an independent field since it is determined by \eqref{constraint}). However, there is no longer any singularity in the moduli space, since there is no point on the moduli space where all the derivatives of the constraint \eqref{constraint} vanish. So, in this case we find (as in similar cases, like the $U(N_c)$ theory with $N_f=N_c$) that the quantum moduli space is a smoothed out version of the classical moduli space.

 In the special case of $N_c=3$ and $N_f=1$, we can compare this with known results, since in this case the flavor is in the adjoint representation, so we have an enhanced $\mathcal{N}=4$ supersymmetry (and our theory is simply the pure ${\cal N}=4$ SYM theory). This case was in fact studied (for an $SU(2) \simeq SO(3)$ gauge group) in \cite{Seiberg:1996nz}.
Due to the $\mathcal{N}=4$ supersymmetry, the moduli space of the $N_c=3$ theory with $N_f=1$ must be a four dimensional hyper-K\"ahler manifold. The set of possible such manifolds (with at most isolated singularities) is very restricted. Using zero-mode counting the authors of \cite{Seiberg:1996nz} were able to determine the precise hyper-K\"ahler manifold for the $SU(2)$ case. It is called the Atiyah-Hitchin manifold $\mathcal{N}$, and it may be defined by the complex surface \cite{Atiyah:1988jp}
\begin{equation}
A^2 = B^2 V +1,
\end{equation}
along with the identification $(A,B,V)\sim(-A,-B,V)$. The similarity of this defining equation with the constraint \eqref{constraint} we found (including now just a single field $M$) is quite satisfying. Presumably the small difference between the two cases (in our case we only have the ${\tilde q} \to -{\tilde q}$ identification) is related to the fact that we consider an $O(3)$ gauge theory while \cite{Seiberg:1996nz} considered an $SU(2)$ gauge theory.

Finally, when $N_f < N_c - 2$, at generic points on the Higgs branch the gauge group is broken to a non-Abelian $O(N_c-N_f)$ gauge theory with no massless flavors, and we know that this theory has no stable supersymmetric vacuum. So, we expect to have no vacuum in this case. The only
effective superpotential allowed by the symmetry charges is
\begin{equation}
W = \left( Y^2 \det(M)\right)^{-\frac{1}{N_c-N_f-2}},
\end{equation}
and by flowing from the previous cases we can see that it is indeed generated in the quantum theory.
Thus, for $N_f < N_c-2$ there is a runaway potential and no supersymmetric vacua exist, as required for the consistency of our duality.

\section{Flows involving the dualities with a Chern-Simons term}

In this section we discuss the relation between the duality we proposed in section \ref{duality}, and the duality for $O(N_c)$ SQCD theories with a non-zero Chern-Simons term that was recently proposed in \cite{Kapustin:2011gh} (generalizing the $U(N_c)$ duality suggested in \cite{Giveon:2008zn}). More generally, we will discuss the consistency of the two dualities with real mass deformations, which allow us to flow between theories with a different Chern-Simons level. We show that in general, when such a deformation is performed, the moduli space is composed of disconnected sets of vacua. Each of these disconnected parts of the moduli space gives a different realization of the duality, and can be matched with the dual theory.

There is a significant simplification in the theory with a Chern-Simons term, since the gauge fields are massive (already classically), and the Coulomb branch is completely lifted. In fact, with a Chern-Simons term, the structure of the moduli space, which contains only the Higgs branch, is quite similar to the moduli space of $d=4$ SQCD. Nevertheless, the possibility of adding real masses is unique to $d=3$, and is interesting to examine closely. The fact that the gauge fields are massive means that whenever the Chern-Simons level $k$ is non-zero, the low-energy dynamics involves only the chiral superfields, and it can be reliably analyzed semi-classically.

Let us begin by reviewing the duality of \cite{Kapustin:2011gh}. The original theory is an $O(N_c)$ gauge theory with a Chern-Simons term of level $k$ (we denote this by $O(N_c)_k$), and with $N_f$ chiral superfields $Q$ in the vector representation. The dual theory is a gauge theory $O(\tilde{N}_c)_{-k}$, where $\tilde{N}_c = N_f + |k| - N_c + 2$, with $N_f$ chiral superfields $q$ in the vector representation, and with extra meson fields $M$ as in the $k=0$ case (with the same superpotential \eqref{generalW} involving the mesons, but without the field $Y$).

As we discussed above, the absence of stable vacua for $N_f < N_c - 2$ is crucial for the consistency of the duality of section \ref{duality}. An analogous assertion holds also for $k \neq 0$. It is known from \cite{Witten:1999ds} that there are supersymmetric vacua, for a pure $SO(N_c)_k$ theory, only provided that $|k| \geq N_c - 2$. Combining this with the result of the previous section we are led to expect that, for an $O(N_c)_k$ theory with $N_f$ chiral superfields, there are supersymmetric vacua only when $N_f + |k| - N_c +2 \geq 0$. This assumption is consistent with the various flows between different theories (by real masses, complex masses and Higgsing).

We begin by comparing the low-energy theories at generic points in the moduli space for the two dual theories, when $N_f+|k| \geq N_c-2$.
For the original $O(N_c)$ theory the classical scalar potential is given by
\begin{equation}
V = g^2 \left( \frac{k}{16 \pi} \phi^a  + Q^\dagger T^a Q \right)^2 + |\phi Q|^2,
\end{equation}
which has zeros only for $\phi=0$. If $N_f \leq N_c$ then at a generic point of the moduli space the gauge group is broken to $O(N_c-N_f)_k$ with no flavors, and the low-energy effective theory contains in addition $\frac{1}{2}N_f(N_f+1)$ massless chiral superfields (abbreviated csf). On the other hand, if $N_f > N_c$ the gauge group is generically completely broken and there are $[N_c N_f - \frac{1}{2}N_c(N_c -1)]$ massless csf.

Let's compare this with the dual theory. The scalar potential is now
\begin{equation}
V = {\tilde g}^2 \left( -\frac{k}{16 \pi} {\tilde \phi}^a  + q^\dagger T^a q \right)^2 + |{\tilde \phi} q|^2 + |qM|^2 + \frac{1}{4}|qq|^2.
\end{equation}
The vacua are still all at $\tilde \phi = 0$.
If $N_f \leq N_c$, all the $\frac{1}{2}N_f(N_f+1)$ components of $M$ are free to condense, and hence all the $q$'s are massive at generic points on the moduli space. The low-energy effective theory is thus an $O(\tilde{N}_c)_{-k}$ theory with no flavors, together with the meson fields. If $N_f > N_c$ then we must have that $\mathrm{Rank}(M) \leq N_c$. Otherwise, we would remain with only $\tilde{N}_f = N_f - \mathrm{Rank}(M)$ flavors such that $\tilde{N}_f + k - \tilde{N}_c + 2 < 0$ and, by the discussion above, there would be no supersymmetric vacua. Therefore $\mathrm{Rank}(M) \leq N_c$, which means that there are $[N_c N_f - \frac{1}{2}N_c(N_c -1)]$ free components in $M$, and in addition at generic points in the moduli space we have a $O(\tilde{N}_c)_{-k}$ gauge theory with $N_f-N_c$ massless flavors.

In summary, at a generic point on the moduli space the low-energy effective theory includes (csf = chiral superfields)

\begin{table}[h]
\begin{center}
\begin{tabular}{|c|p{7cm}|p{8cm}|}
   \hline
            &  original theory  & dual theory \tabularnewline
   \hline
   & & \tabularnewline
$N_f \leq N_c$         & $O(N_c-N_f)_k$, no flavors + $\frac{1}{2}N_f(N_f+1)$ massless csf          & $O(\tilde{N}_c)_{-k}$, no flavors  + $\frac{1}{2}N_f(N_f+1)$ massless csf \tabularnewline & & \tabularnewline
   \hline
   & & \tabularnewline
$N_f > N_c$ & $[N_c N_f - \frac{1}{2}N_c(N_c -1)]$ massless csf   & $O(\tilde{N}_c)_{-k}$, $(N_f-N_c)$ flavors + \ \ \ \ \ \ \ \ \ \ \ \ \ \ \  \ $[N_c N_f - \frac{1}{2}N_c(N_c -1)]$ massless csf \tabularnewline & & \tabularnewline
   \hline
\end{tabular}
\end{center}
\end{table}

\noindent The effective theories we find are consistent with the duality, at least in terms of counting superfields and relating the unbroken gauge theories; for $N_f \leq N_c$ this follows from the relation $k-(N_c-N_f)+2={\tilde N}_c$, and for $N_f > N_c$ because $N_f-N_c+k-{\tilde N}_c+2=0$.

Consider now deforming the original theory by real masses for the quarks. In the original theory this is given by equation \eqref{realmass}, leading to the following scalar potential
\begin{equation}
V = g^2 \left( \frac{k}{16 \pi} \phi^a  + Q^\dagger T^a Q \right)^2 + |(\phi - m_R)Q|^2.
\label{realmass potential}
\end{equation}
To map this deformation to the dual theory we remember the interpretation of the real masses as components of background gauge fields for the $SU(N_f)\times U(1)_A$ flavor symmetry. The mass matrix $m_R$ is in the adjoint of the flavor group, and it couples to the fields of the dual theory charged under this symmetry, $q$ and $M$. This leads to the scalar potential
\begin{equation}
V = {\tilde g}^2 \left(- \frac{k}{16 \pi} {\tilde \phi}^a  + q^\dagger T^a q \right)^2 + |({\tilde \phi} + m_R)q|^2 +  |qM|^2 + \frac{1}{4}|qq|^2 + 4 |[m_R,M]|^2.
\end{equation}

The possible real mass deformations are characterized by the number of positive and negative eigenvalues of the mass matrix $m_R$. If the number of positive eigenvalues of $m_R$ is equal to the number of the negative eigenvalues, there is no net shift in the Chern-Simons level of the low-energy gauge theory. We call this an opposite sign deformation. If the signs are not equal -- a same sign deformation -- there will be a change in the Chern-Simons level according to $k\rightarrow k +\sum \mathrm{sign}[m_\Psi]$, where $\Psi$ represents all massive fermions. We will consider one example of each possibility, the generalization to other cases is straightforward. More examples of real mass flows, both for orthogonal and for other classical gauge groups, are analyzed in detail in \cite{ItamarThesis}.

Consider first an opposite sign deformation of some $k\neq 0$ theory, which does not change the level. We take the number of flavors to be even in this case, and consider a real mass matrix of the form
\begin{equation}
m_R = m \sigma^3 \otimes \mathbf{1}_{\frac{1}{2}N_f \times \frac{1}{2} N_f}.
\end{equation}
We will now find the set of vacua on both sides by finding configurations with a vanishing scalar potential, and showing that they match.

In the original theory the solutions take the form
\begin{equation}
\phi =  \sigma^3 \otimes
\begin{pmatrix}
  m \mathbf{1}_{L } &  \\
 					        & 0 &  \\
   					        &   & \ddots & \\
  					        &   &        & 0
 \end{pmatrix}.
\end{equation}
For each $L$ the symmetry breaking pattern is $O(N_c) \rightarrow U(L) \times O(N_c - 2L)$ \footnote{The appearance of the unitary group here is equivalent in the standard representation of the orthogonal group $O(2n)$ to the well-known statement that the set of orthogonal $2n\times 2n$ matrices that commutes with $\sigma^2 \otimes \mathbf{1}_{n \times n}$ is isomorphic to $U(n)$.}. From the upper block, we have a $U(L)_k$ theory with $\frac{1}{2} N_f$ csf in the $\mathbf{L}$ representation and $\frac{1}{2} N_f$ in the $\overline{\mathbf{L}}$ representation.  Solving the restriction of \eqref{realmass potential} to the upper block we find that the real mass effectively behaves as a Fayet-Iliopoulos term in the $U(L)$ block, forcing the flavors to acquire VEVs leading to a complete breaking of $U(L)$. One can show that there are vacua only when $\frac{1}{2}N_f \geq L$, and that in this case there are $N_f L - L^2$ massless csf left at generic points on the moduli space. In the lower block there is an $O(N_c - 2L)_k$ theory with no massless flavors.

In the dual theory, taking similarly ${\tilde \phi} = \sigma^3 \otimes \mathrm{diag}(m\mathbf{1}_P,0,\ldots,0)$, similar considerations lead to the conclusion that there are vacua only when $\frac{1}{2}N_f \geq P$, in which case there are $\frac{1}{2}N_f P-P^2$ massless csf. This is different than above since, in the dual theory, the superpotential does not allow the flavors in the $\mathbf{P}$ of $U(P)$ and those in the $\overline{\mathbf{P}}$ to simultaneously condense, though the effective Fayet-Iliopoulos term forces one of them to condense. In addition, there is generically an $O(\tilde{N}_c)_{-k}$ theory with no flavors, and $\frac{1}{4}N_f^2 - \frac{1}{2}N_f P$ massless csf coming from $M$.

There is a one-to-one correspondence between the set of vacua of both sides. The mapping is given by the relation $L+P=\frac{1}{2} N_f$, such that for each $L=0,\ldots,\frac{1}{2} N_f$ we have the two dual gauge theories
\begin{equation}
O(N_c-2L)_k \simeq O(\tilde{N}_c + 2L - N_f)_{-k},
\end{equation}
and in addition $N_f L - L^2 = \frac{1}{4} N_f^2 - P^2$ massless csf.

Consider now a same sign deformation, in which there is a net change of the Chern-Simons level. As an example we give a mass to only a single flavor ($N_f$ is not constrained)
\begin{equation}
m_R = \mathrm{diag}(0,\ldots , 0 , m).
\end{equation}
The low-energy Chern-Simons level in this case is shifted by one, and the semi-classical analysis is valid whenever the low-energy Chern-Simons level is non-zero.
Using the same arguments as above, we may find the vacuum structure on the two sides. The possible vacua depend on the sign of $m$. When $m>0$ the only solutions
are given by
\begin{table}[h]
\begin{center}
\begin{tabular}{|c|c|}
   \hline
original theory  & dual theory \tabularnewline
   \hline
   & \tabularnewline
        $O(N_c)_{k+1}$ with $N_f-1$ flavors & $O(\tilde{N}_c)_{-k-1}$ with $N_f-1$ flavors\tabularnewline
   & \tabularnewline
   \hline

\end{tabular}
\end{center}
\end{table}

\noindent with no breaking of the gauge symmetry. The two low-energy theories are still dual to each other, after the
shift to the Chern-Simons level on both sides coming from the massive charged fields that have been integrated out.

Next, when $m<0$, there are two vacua on each side (with different VEVs for $\phi$, $\tilde \phi$), whose low-energy descriptions (which are still dual to each other) are given by
\begin{table}[h]
\begin{center}
\begin{tabular}{|c|c|}
   \hline
original theory  & dual theory \tabularnewline
   \hline
   & \tabularnewline
        $O(N_c)_{k-1}$ with $N_f-1$ flavors & $O(\tilde{N}_c-2)_{-k+1}$ with $N_f-1$ flavors\tabularnewline
   & \tabularnewline
   \hline
   & \tabularnewline
        $O(N_c-2)_{k-1}$ with $N_f-1$ flavors & $O(\tilde{N}_c)_{-k+1}$ with $N_f-1$ flavors\tabularnewline
   & \tabularnewline
   \hline
\end{tabular}
\end{center}
\end{table}

 The same-sign real mass deformation can be analyzed in this way even if we start from the duality of section \ref{duality} with $k=0$, since the low-energy theory in this case has non-zero $k$; in fact
 this gives a flow between our duality and that of \cite{Kapustin:2011gh}.
 In particular, such a real mass deformation, which includes a background field of the $U(1)_A$ flavor symmetry under which also $Y$ and $\tilde{Y}$ are charged, means that these operators receive real masses, and play no role in the duality with the Chern-Simons term.

Analyzing real mass flows such that there is no Chern-Simons level in the low-energy theory is much
more complicated, since one cannot trust the semi-classical analysis, and we leave it to future work.
In particular one can flow also in the other direction, starting with a theory with non-zero Chern-Simons
level and ending up with no Chern-Simons level at low energies, but it is not clear how the Coulomb
branch variables $Y$ and $\tilde Y$ show up in such a flow (presumably there should be some way to incorporate these fields into the duality with non-zero $k$).

\section{Summary}

In this paper we analyzed the low-energy dynamics of $d=3$ ${\cal N}=2$ supersymmetric $O(N_c)$ SQCD theories. We found a picture very similar to the one found for other classical gauge groups; for $N_f < N_c-2$ there is no stable vacuum, for $N_f = N_c-2$ there is a smooth moduli space, while for $N_f > N_c-2$ there is an interacting fixed point at the origin of the moduli space, which has a dual description in terms of an $O(N_f-N_c+2)$ gauge theory. The main difference between the $O(N_c)$ case and other cases is that the consistency of these results required using as a coordinate on the moduli space the chiral superfield $Y$ corresponding to the minimal magnetic charge, even though it does not correspond to a semi-classical 't~Hooft-Polyakov monopole, while for other classical gauge groups the correct coordinate does arise from a semi-classical monopole.  We provided various consistency checks for this choice of coordinate, but it would be nice to understand it better. We showed that in some cases there was a nice relation between the duality for $O(2)$ and $SO(2) \simeq U(1)$ groups, while in other cases these two dualities lead to different theories (which should be related by a charge conjugation identification). It would be interesting to try to find a nice effective description for general $SO(N_c)$ gauge theories, which (as for $SU(N_c)$ theories) is not yet available.

We hope that our results could be useful for a more systematic understanding (and perhaps a derivation) of Seiberg dualities. Specifically it would be nice to understand the flow from the dualities with a non-zero Chern-Simons level \cite{Giveon:2008zn,Kapustin:2011gh} to the theory with vanishing Chern-Simons level, that can be realized by giving real masses to some of the flavors. Since the effective description of the theories with Chern-Simons couplings does not include the Coulomb branch coordinate $Y$, understanding such a flow requires understanding the role of $Y$ in the theory with Chern-Simons couplings, and of how it couples to the other fields in its effective description.

It has recently been suggested that renormalization group flows in $d=3$ theories (and, in particular, in $d=3$ ${\cal N}=2$ supersymmetric theories) obey a $Z$-theorem such that the logarithm of their partition function on the three-sphere always decreases \cite{Jafferis:2010un}. We hope that the new fixed points and renormalization group flows described in this paper can be useful for checking this suggestion.

Finally, it would be interesting to understand better the dynamics of $d=4$ $O(N_c)$ SQCD theories compactified on a circle; in the small radius limit they should reproduce the dynamics we described here. We discussed the compactified theories here just in the pure SYM case, but one can try to discuss them also for $N_f > 0$, as done for the $U(N_c)$ and $USp(2N_c)$ cases \cite{Aharony:1997bx,Aharony:1997gp}. One problem that arises only in the $O(N_c)$ case is that the superpotential related to the circle compactification is proportional to a different chiral superfield ($Z$) than the one ($Y$) which we found useful for describing the effective dynamics of the $d=3$ quantum theory. Thus, it is not clear what are the useful variables for an effective description of the $d=4$ theory on a circle, and how to write down a superpotential for this theory (the partition function identities of \cite{Dolan:2011rp} could provide useful clues for this). Another issue is that the Seiberg-dual gauge group in $d=4$, $O(N_f-N_c+4)$, is larger than the $d=3$ dual group $O(N_f-N_c+2)$, suggesting that in the dual description the gauge group is partially broken by the circle compactification; again it is not clear how this arises in the effective description of the $d=4$ theory on a circle.

%~~~~~~~~~~~~~~~~~~~~~~~~~~~~~~~~~~~~~~~~~~~~~~~
\subsection*{Acknowledgments}
\label{s:acks}
%~~~~~~~~~~~~~~~~~~~~~~~~~~~~~~~~~~~~~~~~~~~~~~

We would like to thank F. Benini, J. Erlich, A. Giveon, D. Jafferis, Z. Komargodski, D. Kutasov, I. Yaakov, and especially C. Closset and S. Cremonesi for useful discussions.
This work was supported in part by the Israel--U.S.~Binational Science Foundation, by a research center supported by the Israel Science Foundation (grant number 1468/06), by the German-Israeli Foundation (GIF) for Scientific Research and Development, and by the Minerva foundation with funding from the Federal German Ministry for Education and Research.

\appendix

\section{$SO(N)$ conventions}\label{appA}

We will use the following representation for the orthogonal group \cite{Sepanski:2006}. Define
\begin{align}
T_{2n}  = & \frac{1}{\sqrt{2}}
        \left(\begin{array}{cc}
			\mathbf{1}_{n} & \mathbf{1}_{n}\\
			i\mathbf{1}_{n} & -i\mathbf{1}_{n}\end{array}\right), &
E_{2n}= & \left(\begin{array}{cc}
			0 & \mathbf{1}_{n}\\
			\mathbf{1}_{n} & 0\end{array}\right),\\
T_{2n+1} = & \mathrm{diag}(T_{2n},1), & E_{2n+1} =&  \mathrm{diag}(E_{2n},1),						
\end{align}
\begin{align}
SO(E_{N}) = & \{g\in SL(N,\mathbb{C})| \quad g^*=E_{N}gE_{N}, \quad g^{T}E_{N}g=E_{N}\},\\
\mathfrak{so}(E_{N})  = & \{X\in\mathfrak{gl}(N,\mathbb{C})| \quad X^*=E_{N}XE_{N}, \quad X^{T}E_{N}+E_{N}X=0\}.
\end{align}
The map $g\rightarrow T_{N}^{-1}gT_{N}$ is a Lie group isomorphism that takes the usual representation of $SO(N)$ by orthogonal matrices to $SO(E_{N})$, and an induced isomorphism of the Lie algebras follows. The advantage of this representation is that the matrices in the Cartan subalgebra are diagonal :
\begin{align}
\mathfrak{t}_{2n} =  (\theta_{1},\ldots,\theta_{n},-\theta_{1},\ldots,-\theta_{n}), \quad \quad
\mathfrak{t}_{2n+1} =  (\theta_{1},\ldots,\theta_{n},-\theta_{1},\ldots,-\theta_{n},0).
\end{align}
For $N=2n$ the Weyl group acts by all permutations of $\theta_i$ and all changes of an even number of signs of $\theta_i$. For the group $O(2n)$ all sign changes are allowed. For $N=2n+1$ the Weyl group acts by all permutations and all sign changes in both cases.

The dual simple roots of $SO(E_{2n})$ are given by ($1 \le i \le n-1$)
\begin{align}
h_i  =  (e_{i}-e_{i+1})-(e_{i+n}-e_{i+n+1}), \quad \quad
h_{n}  =  (e_{n-1}+e_{n})-(e_{2n-1}+e_{2n}),
\end{align}
where $e_{i}=\mathrm{diag}(0,\ldots0,1,0,\ldots,0)$ with $1$ in the $i^{th}$ position. For $SO(E_{2n+1})$ they are
\begin{align}
h_i  = (e_{i}-e_{i+1})-(e_{i+n}-e_{i+n+1}), \quad \quad
h_{n}  = 2e_{n}-2e_{2n},
\end{align}
with $1 \le i \le n-1$.

\section{Charge quantization of magnetic monopoles}
\label{appB}

The Dirac quantization of magnetic monopole charge can be understood by considering all possible $U(1)$ bundles over a sphere $S^2$. We can work with two patches, one containing the north pole $U_N$ and one the south pole $U_S$, in the standard way. The possible bundles are classified by the set all possible transition functions, which are maps from the intersection $U_N \cap U_S$, which is essentially the equator ($\simeq S^1$), to the gauge group $U(1)$. This means that the allowed bundles are classified by the homotopy group $\pi_1(U(1))=\mathbb{Z}$. A different way of saying the same thing is that the transition functions $t_{NS}: S^1 \rightarrow U(1)$ are of the form $\exp(i \theta H)$, with $0 \le \theta \le 2\pi$, and that for $t_{NS}$ to be well defined we must have that $\exp(2\pi i H)=1$, which defines a lattice in the Lie algebra of $U(1)$.

We want to consider this quantization in a generalized setting. We take a general Abelian group $U(1)^n$, which is the maximal torus $H$ of a semi-simple Lie group $G$. Repeating the above argument means that the possible bundles are in one to one correspondence with the solutions of $\exp(2 \pi i H) = I$, with $H$ in the Cartan algebra $\mathfrak{t}$. This defines the following lattice
\begin{equation}
\mathrm{ker} \,\mathcal{E} = \{ H \in \mathfrak{t} | \exp(2 \pi i H) = I \}.
\end{equation}

A useful theorem gives the relation $\pi_1(G) \cong \mathrm{ker} \, \mathcal{E}/R^{\vee}$, where $R^{\vee}$ is the dual root lattice \cite{Sepanski:2006}. Then, for a simply connected group (such as $SU(N)$ or $USp(2N)$) the magnetic charges are given by the dual roots. For $SO(N)$ we may use another theorem stating that $Z(G) \cong P^{\vee}/ \mathrm{ker} \, \mathcal{E}$, where $Z(G)$ is the center of $G$ and $P^{\vee}$ is the dual weight lattice \cite{Sepanski:2006}. This means, for example, that there are more solutions to the quantization conditions in $SO(3)$ then there are in $SU(2)$, since $R^{\vee} \subseteq P^{\vee}$.

We can ask whether all the allowed magnetic charges correspond to 't~Hooft-Polyakov monopoles. To answer that we can reason as follows. In the 't~Hooft-Polyakov setting we have a $G$ bundle over $\mathbb{R}^3$ (which is necessarily trivial), whose structure group is reduced to $H$ on the sphere $S^2$. We may take two patches for $\mathbb{R}^3$ intersecting at the $x-y$ plane. The transition functions are now defined on the disk $D_2$ whose boundary is the equator ($\simeq S^1$) from before. The transition function therefore defines a homotopy of the path $\exp(i\theta H)$ to a trivial path at the origin. This means that the element of $\pi_1(H)$ which correspond to a 't~Hooft-Polyakov monopole must be trivial when embedded in $G$, i.e. it sits in the dual root lattice $R^{\vee}$ (by the theorem above). In particular for $SO(N)$ the minimal magnetic charges do not correspond to 't~Hooft-Polyakov magnetic monopoles.

\end{document}